\pdfoutput=1
\RequirePackage[2020-02-02]{latexrelease}
\documentclass[a4paper,american,citeautoscript,floatfix,pdftex,superscriptaddress,twoside,%
aip,jcp,%
reprint%,final% add final to see real layout, no todonotes anymore, ...
]{revtex4-2}%
\usepackage{amsfonts,amsmath,amssymb}
\usepackage{booktabs}
\usepackage{commath}
\usepackage{dcolumn}
\usepackage[T1]{fontenc}
\usepackage{graphicx}%
\usepackage[utf8]{inputenc}
\usepackage{microtype}
\usepackage{hyperref, hypernat}
\usepackage[todos]{CMImacros}

\usepackage{siunitx}

\graphicspath{{./figures/}} % define figure directory
\hypersetup{citebordercolor=yellow,linkbordercolor=red,urlbordercolor=blue} %

\newcommand{\cfeldesy}{\affiliation{Center for Free-Electron Laser Science CFEL, Deutsches
      Elektronen-Synchrotron DESY, Notkestraße 85, 22607~Hamburg, Germany}}%
\newcommand{\uhhcui}{\affiliation{Center for Ultrafast Imaging, Universität Hamburg, Luruper
      Chaussee 149, 22761 Hamburg, Germany}}%
\newcommand{\uhhphys}{\affiliation{Department of Physics, Universität Hamburg, Luruper Chaussee 149, 22761 Hamburg, Germany}}%
\newcommand{\desy}{\affiliation{Deutsches Elektronen-Synchrotron DESY, Notkestraße 85, 22607 Hamburg, Germany}}%
\newcommand{\jena}{\affiliation{Helmholtz-Institute Jena, Fröbelstieg 3, 07743 Jena, Germany}}%
\newcommand{\gsi}{\affiliation{GSI Helmholtzzentrum für Schwerionenforschnung GmbH, Planckstrasse 1, 64291 Darmstadt, Germany}}%
\newcommand{\ihemail}{\email[Email:~]{ingmar.hartl@desy.de}}%
\newcommand{\jkemail}{\email[Email:~]{jochen.kuepper@cfel.de}}%
\newcommand{\cmiweb}{\homepage[website:~]{https://www.controlled-molecule-imaging.org}}%

\begin{document}
\title{Self-broadening and self-shift in the $\mathbf{3\nu_{2}}$ band of ammonia from
   mid-infrared-frequency-comb spectroscopy }%
\author{Guang Yang}\cfeldesy\uhhphys%
\author{Vinicius Silva de Oliveira}\uhhphys\desy%
\author{Dominic Laumer}\uhhphys\desy%
\author{Christoph M. Heyl}\desy\jena\gsi
\author{Andrey Yachmenev}\cfeldesy\uhhcui%
\author{Ingmar Hartl}\ihemail\desy%
\author{Jochen Küpper}\jkemail\cmiweb\cfeldesy\uhhphys\uhhcui%
\begin{abstract}\noindent%
   We report the broadband absorption spectrum of the $3\nu_{2}$ band of $^{14}$NH$_{3}$ near 4~\um. The data were
   recorded using a mid-infrared frequency comb coupled to a homebuilt Fourier-transform spectrometer with a resolution of
   0.00501~\invcm. Line positions, line intensities, self-broadening, and self-shift parameters for six rovibrational lines were
   determined at room temperature ($T=296$~K). Comparison with HITRAN 2016 shows good agreement at improved precision.
   The high precision and the rapid tunability of our experiment enables advanced fast spectroscopy of molecular gases.
\end{abstract}
\maketitle

\section{Introduction}
\label{sec:introduction}
Ammonia is one of the spectroscopically most-studied molecules due to its importance in the atmosphere, astrophysics,
chemistry, medicine and biology. Ammonia exists in a wide range of astrophysical environments and was the first polyatomic
molecule detected in the interstellar medium~\cite{Cheung:PRL21:1701}. It was used as one of the most accurate ``molecular
thermometers'' to measure the temperature of C/2001 Q4(Neat)~\cite{Kawakita:AJ643:1337},
9P/Tempel~1~\cite{Kawakita:DI191:513} and other comets by detecting the \textit{ortho-para} ratio. Ammonia also formed the
basis for modern laser technology through the original demonstration of the MASER~\cite{Gordon:PR99:1264}. Ammonia spectra
have been extensively studied and disentangled, both in theory and experiment.

The spectra of ammonia were extensively measured from the microwave to the ultraviolet~\cite{Gordon:PR95:282,
Urban:JMS88:274, Drouin:JMS1006:2, Hillman:AO18:1808, Cottaz:JMS209:30, Kleiner:JMS173:120, Pearson:JMS353:60,
Li:JMS243:219, Langford:JCP108:6667, Douglas:DFS35:158, Veldhoven:EPJD31:337}. Yurchenko
\etal~\cite{Yurchenko:JPCA113:11845} calculated the spectra for ammonia covering a large part of the infrared region using a
variational approach. Recently, we calculated hyperfine-resolved rotation-vibration line lists of
ammonia~\cite{Yachmenev:JCP147:141101, Coles:ApJ870:24}, which agree very well with experimental
results~\cite{Twagirayezu:JCP145:144302}. The rotation-vibration-spectroscopy data were all critically reviewed and included in
the ``Measured Active Rotational-Vibrational Energy Levels (MARVEL)'' database~\cite{Alderzi:JQSRT161:117}. The
high-resolution transmission (HITRAN) molecular absorption database~\cite{Gordon:JQSRT203:3} summarizes the ammonia
rotation-vibration spectra, but still has many of the transitions unassigned or absent, especially in the molecular fingerprint
region in the mid-infrared (MIR). For instance, there are only two experiments in the 4~\um region, obtained using a Globar
source~\cite{Kleiner:JMS173:120} and a synchrotron source~\cite{Pearson:JMS353:60}.

The development of frequency combs (FC) extended the traditional gas-phase-absorption spectroscopy to broadband, which
allows for the simultaneous detection of multiple transition regions of multiple species in a short acquisition
time~\cite{Cossel:JOSAB34:104}. Coherent MIR-FC light sources allow to precisely measure molecular fingerprints which are
useful, \eg, in breath analysis, atmospheric measurements, and astrochemistry~\cite{Cossel:JOSAB34:104}. Taking breath
analysis for an example, ammonia sensing in exhaled human breath could be an indicator of liver and renal
diseases~\cite{Sigrist:GasSensing}. Nonlinear frequency conversion approaches utilizing optical parametric oscillators
(OPO)~\cite{Tillman:APL85:3366, Kornaszewski:OE15:11219, Adler:OE18:21861}, optical parametric amplification
(OPA)~\cite{Heiner:OE26:25793}, supercontinuum generation~\cite{Cheng:OL16:2117, Zhao:OL41:5222}, or difference frequency
generation (DFG)~\cite{Ruehl:OL12:2232, Sobon:OL42:1748} provide an alternative approach to generate optical frequencies in
MIR regions which are not covered by laser gain media. MIR FC sources based on DFG are widely used due to their simplicity,
single-pass configuration, broad tunability, and full cancellation of the carrier-envelop offset. The combination of a MIR FC with a
Fourier transform spectrometer (FTS) offers capabilities to record spectra over a broadband range and with high resolution, high
sensitivity, and especially a high signal-to-noise ratio (SNR) due to the high spectral brightness and temporal coherence of
FCs~\cite{ Tillman:APL85:3366, Kornaszewski:OE15:11219, Adler:OE18:21861}.

Here, we report the absorption spectrum of the $3\nu_{2}$ band of $^{14}${NH}$_{3}$ near 4~\um by a homebuilt FTS coupled
to a DFG-based MIR FC covering the range 3--6~\um. We extract the transition wavenumbers and intensities for 6 typical $R$
branch rovibrational lines through a multipeak Voigt fit at room temperature (296~K). We also retrieve self-broadening- and
self-shift-parameters of the pressure interactions in the gas at nine different pressures ranging from 10.00~mbar to
700.00~mbar.

\section{Experiment}
\label{sec:experiment}
\begin{figure}
   \includegraphics[width=\linewidth]{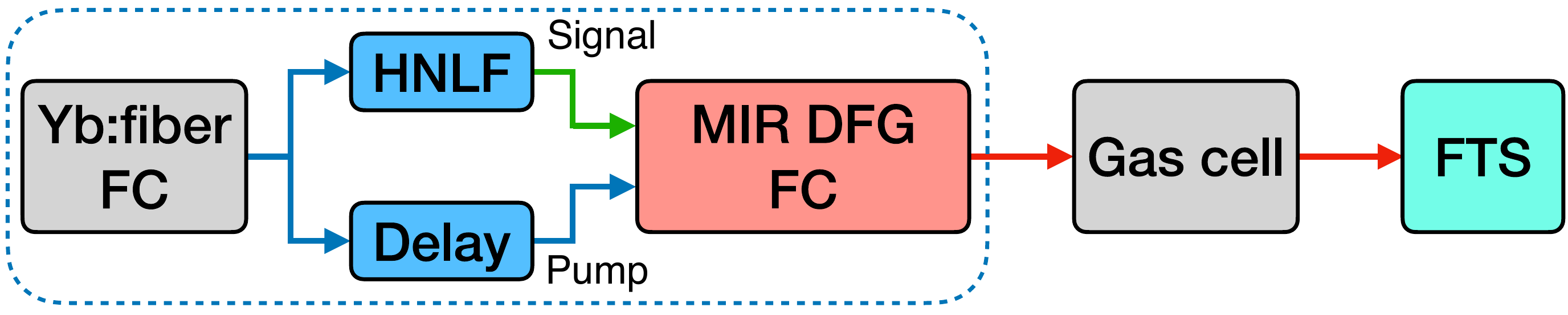}%
   \caption{Generic schematic of experimental setup. FC: frequency comb, HNLF: highly-nonlinear suspended core-fiber, MIR DFG
   FC: DFG-based mid-infrared FC, FTS: Fourier transform spectrometer.}
   \label{fig:setup}
\end{figure}
A schematic diagram of the experimental setup is shown in \autoref{fig:setup}. It consists of three parts: the DFG-based MIR FC
source, the gas cell, and the FTS. The MIR FC source consists of a Yb:fiber FC and a DFG setup~\cite{Ruehl:OL12:2232,
SilvadeOliveira:OL7:1914}. The Yb:fiber FC is mode-locked by a saturable absorber with the repetition rate of 150~MHz and
phase-locked to a cw laser (Coherent Mephisto) with kHz-level linewidth at $\ordsim1064$~nm and frequency stability to
1~MHz/minute. Using chirped-pulse amplification, 1.5~W average power is deliverd in pulses of 130~fs duration with a center
wavelength of 1050~nm and a bandwidth of 25~nm. The Yb:fiber FC is split into pump and signal driver of the DFG process, with
the signal generated as the longest Raman soliton from the supercontinuum process in a highly-nonlinear suspended
core-fiber~\cite{Dong:OE16:16423}. By controlling the input power, we could adjust the wavelength of the longest Raman soliton
from 1.2~\um to 1.6~\um. The signal and pump pulse are temporally overlapped in a fan-out periodically poled Lithium Niobate
(PPLN) crystal (HCPhotonics) for DFG by locking a mechanical delay-line in the pump arm~\cite{SilvadeOliveira:OL7:1914}. This
yielded MIR radiation ranging from 3~\um to 6~\um. The MIR FC was tuned to near 4~\um (2390~\invcm to 2510~\invcm), and the
light passed through the gas cell together with a frequency stabilized (1~MHz/minute) HeNe laser (SIOS SL4), which was used to
calibrate the absolute frequency of the spectroscopic signals.

Ammonia ($^{14}${NH}$_{3}$, Linde HiQ 6.0) was contained in a 35.4646(12)~mm long gas cell with
wedged CaF$_{2}$ windows (30~arcmin) at 296~K. The optical path length was determined using
low-coherence interferometry~\cite{Hartl:OL26:608} utilizing the residual 1050~nm laser light from
the DFG process as the input signal. The optical path difference (OPD) of the four reflected beams
from the two gas cell windows are detected by interfering the reflections with a reference laser.
Before the measurements, the reservoir was pumped (HiCube 80 Eco) to
$\smaller 3.1\times10^{-4}$~mbar, filled to 699.75~mbar of ammonia, and then pumped to 543.85~mbar,
500.01~mbar, 400.50~mbar, 300.70~mbar, 199.90~mbar, 98.90~mbar, 50.00~mbar, and 10.00~mbar,
respectively. All pressures were measured using calibrated ceramic capacitance gauges (Pfeiffer
Vacuum CMR 361 and CMR 364) with a relative accuracy of 0.2\%.

The two laser were both directed to our homebuilt FTS, which is based on a Michelson interferometer with two
liquid-nitrogen-cooled indium antimonide (InSb) detector (InfraRed Associates IRA 20-00060). The interferograms of both MIR FC
and HeNe laser were acquired at 5~MSa/s sampling rate with 20~bit resolution by a computer-controlled data acquisition board
(National Instruments PCI-5922). The interferogram of the HeNe laser was used to measure the OPD and provides steps to
resample the MIR FC interferogram by determining the positions of the zero-crossings. By matching the delay range of the FTS to
the repetation rate of the MIR FC, we overcame the resolution limitation of conventional FTS given by the maximum delay range
and removed the instrumental line shape oscillations, which allowed a reduction of the instrument size and acquisition
time~\cite{Piotr:PRA93:021802}. The grid spacing was equal to the comb spacing of 150~MHz, yielding a resolution of the FTS of
0.00501~\invcm. After performing a fast Fourier transform (FFT) to MIR FC interferograms, we obtained MIR absorption spectra
of ammonia at the different specified pressures. To extract the transmittance, we first measured the reference spectra before
filling ammonia when the reservoir was under vacuum.

\begin{figure}
   \includegraphics[width=\linewidth]{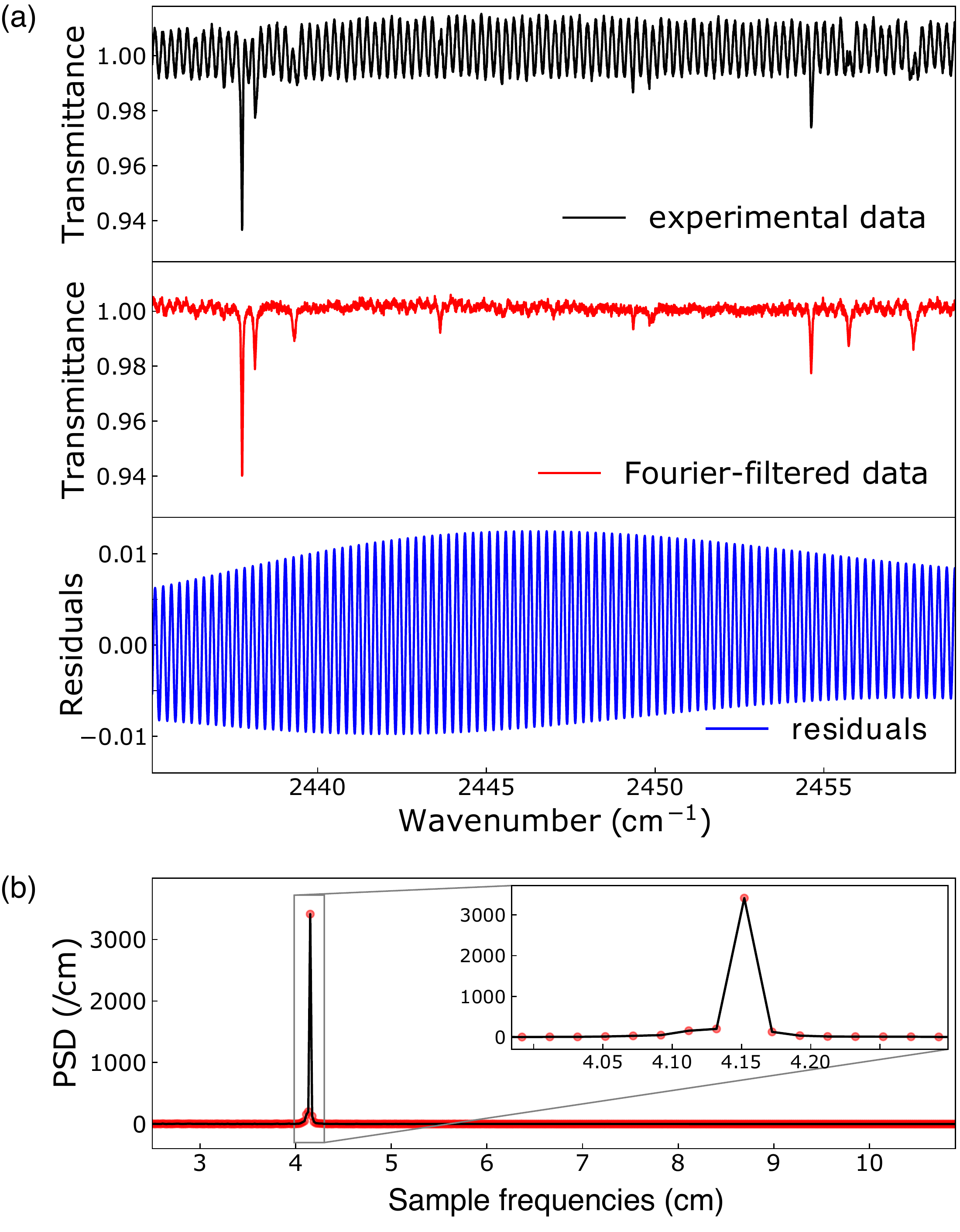}%
   \caption{(a)~The absorption spectra at 98.90 mbar and at room temperature (296~K). The top panel shows the experimentally
   recorded original spectrum. The middle plot shows the spectra after FFT-filtering. The bottom plot shows the residual
   differences of the two upper spectra. (b)~Partial Power Spectral Density (PSD) of the absorption spectrum in (a).}
   \label{fig:spectra}
\end{figure}
The top panel of \autoref[a]{fig:spectra} shows the raw spectrum in the range 2434 to 2459~\invcm at 98.90~mbar and at
296~K. The recorded spectra had oscillatory background signals (black curve in \autoref[a]{fig:spectra}) that not only decreased
the SNR of the spectra, but also changed the center and linewidth of the absorption line profiles. We assigned this regular signal
to etalon effects of the optical components in the experimental setup. In order to remove this etalon signal, we used a numerical
approach based on FFT analysis. Firstly, we calculated the power spectral density (PSD) for different FFT sample frequencies of
the spectra, shown in \autoref[b]{fig:spectra}. The unit of sample frequencies was in cm, which exactly corresponded to the OPD
of the FTS. Then we set the FFT values for the four sample frequencies between 4.10~cm and 4.18~cm to be zero, and performed
an inverse FFT, resulting in the filtered spectrum shown in \autoref[a, middle panel]{fig:spectra}. Analyzing the sample
frequencies of the filtered signal, \autoref[a, bottom panel]{fig:spectra}, we suppose that the etalon signal most likely comes
from an Germanium (Ge) filter in the experimental setup; the length and refractive index of the Ge filter at 4~\um are around
5~mm and 4, resulting in a 4~cm OPD that matches the free spectral range of these oscillations. The FFT filtering might have
introduced some small errors as the filtered frequencies also included real spectral information. Nevertheless, these errors
should be smaller than profile fitting error and the approach clearly improved the precision and SNR. In the future, the Ge filter
shall be replaced by wedged optics.

\section{Results and discussion}
\label{sec:results}
\begin{table*}
   \centering
   \setlength\tabcolsep{0pt}
   \caption{Line transitions, line intensities, self-broadening, and self-shift coefficients for the $3\nu_{2}$ band of
   $^{14}${NH}$_{3}$ at 296 K.}
   \label{tab:results}
   \begin{tabular*}{\textwidth}{@{\extracolsep{\fill}} l*{2}{lS[table-number-alignment = left]llS[table-number-alignment = left]S[table-number-alignment = left]S[table-number-alignment = left]}}
     \toprule
     Transition        & \multicolumn{1}{c}{$\nu_{ij,\,\text{exp.}}^0$} &
     \multicolumn{1}{c}{$\nu_{ij,\,\text{ref.}}$ \footnotemark[1]} & \multicolumn{1}{c}{$I_{ij,\,\text{exp.}}$}
     & \multicolumn{1}{c}{$I_{ij,\,\text{ref.}}$ \footnotemark[1]}
                                                                 &\multicolumn{1}{c}{$
                                                                   \gamma_{\text{self}\,\text{exp.}} $ }
     &\multicolumn{1}{c}{$\gamma_{\text{self}\,\text{ref.}}$ \footnotemark[1]}     &\multicolumn{1}{c}{$\delta_\text{self}$}\\
     (R branch)        & \multicolumn{1}{c}{$(\invcm)$}             & \multicolumn{1}{c}{$(\invcm)$ }  & \multicolumn{1}{c}{$(\text{cm/molec.})$}             & \multicolumn{1}{c}{$(\text{cm/molec.})$ }  &\multicolumn{1}{c}{$(\invcm/atm)$}         &\multicolumn{1}{c}{$(\invcm/atm)$}   &\multicolumn{1}{c}{$(\invcm/atm)$}\\
     \midrule
     $aR(2,0)$         &2437.76258(18)            & 2437.7655    & $5.12(19)\times10^{-22}$ &
     $4.590\times10^{-22}$ &  0.225(13)       &  0.302     & 0.0016(5)\\
     $aR(2,1)$         &2438.14454(106)         & 2438.1475     & $2.35(21)\times10^{-22}$ &
     $2.078\times10^{-22}$ &  0.320(33)       &  0.396     & -0.0112(28)\\
     $aR(2,2)$         &2439.31026(421)         & 2439.3148      & $1.70(20)\times10^{-22}$ &
     $1.372\times10^{-22}$ &  0.445(63)       &  0.496    & -0.0139(110)\\
     $aR(3,1)$         &2454.62688(68)            & 2454.6298     & $2.26(16)\times10^{-22}$ &
     $2.323\times10^{-22}$ &  0.263(31)       &  0.36      & -0.0073(18)\\
     $aR(3,2)$         &2455.73934(231)         & 2455.7392     & $1.93(15)\times10^{-22}$ &
     $1.963\times10^{-22}$ &  0.397(43)       &  0.448     & 0.0006(62)\\
     $aR(3,3)$         &2457.66341(317)         & 2457.6615      & $2.27(21)\times10^{-22}$
     &$2.508\times10^{-22}$ &  0.464(62)       &  0.541    & -0.0351(83)\\
     \bottomrule
   \end{tabular*}
   \footnotetext[1]{Data from the HITRAN~2016 database~[\onlinecite{Gordon:JQSRT203:3, Kleiner:JMS173:120,
   Nemtchinov:JQSRT83:243}], the uncertainty range for $\nu_{0}$, $I_{ij}$ and $\gamma_{\text{self}}$ are 0.0001~\invcm to
   0.001~\invcm, 10\% to 20\%, and 2\% to 5\%, respectively.} %
\end{table*}
Direct multipeak Voigt fit for the transmittance data was performed to retrieve spectral information. Voigt profile was still valid
because the transmittance was more than 90\% for all transitions due to the low sensitivity with the short gas cell and the
pressures of gas sample were relatively high~\cite{Gross:JCP154:054305}. The four basic spectroscopic line parameters -- the
pressure-broadening $\gamma_{p,t}$, the Doppler-broadening $\alpha_{D}$, line transition frequency $\nu_{ij}$, and the
spectral line intensity $I_{ij}$ -- were determined from individual spectra to extract details on the molecular motions and
collisions through a non-linear least-squares fit. Since $\alpha_{D}$ depends on $\nu_{ij}$ and temperature, which was 296~K
here, we could reduce the Voigt fit to the three other parameters, as shown in~\autoref{equ:Voigt}.

\begin{align}
	\label{equ:Voigt}
	T(\nu) = e^{I_{ij}(T)Nl_{path}V(\nu-\nu_{ij},\gamma_{p,t})}+a_0
\end{align}

Here, $N$ is the number density of ammonia, equals to $p/k_{b}T$ assuming the ideal gas law, and $l_{path}$ is the optical path
length (35.4646 mm in our experiment). In addition, small instrumental baseline $a_0$ was added in the fit. We assumed $a_0$
to be a constant value around 0 since the low sensitivity, more general baseline fitting process could be found from
Ref.~\onlinecite{Gross:JCP154:054305}.

\begin{figure}
   \includegraphics[width=\linewidth]{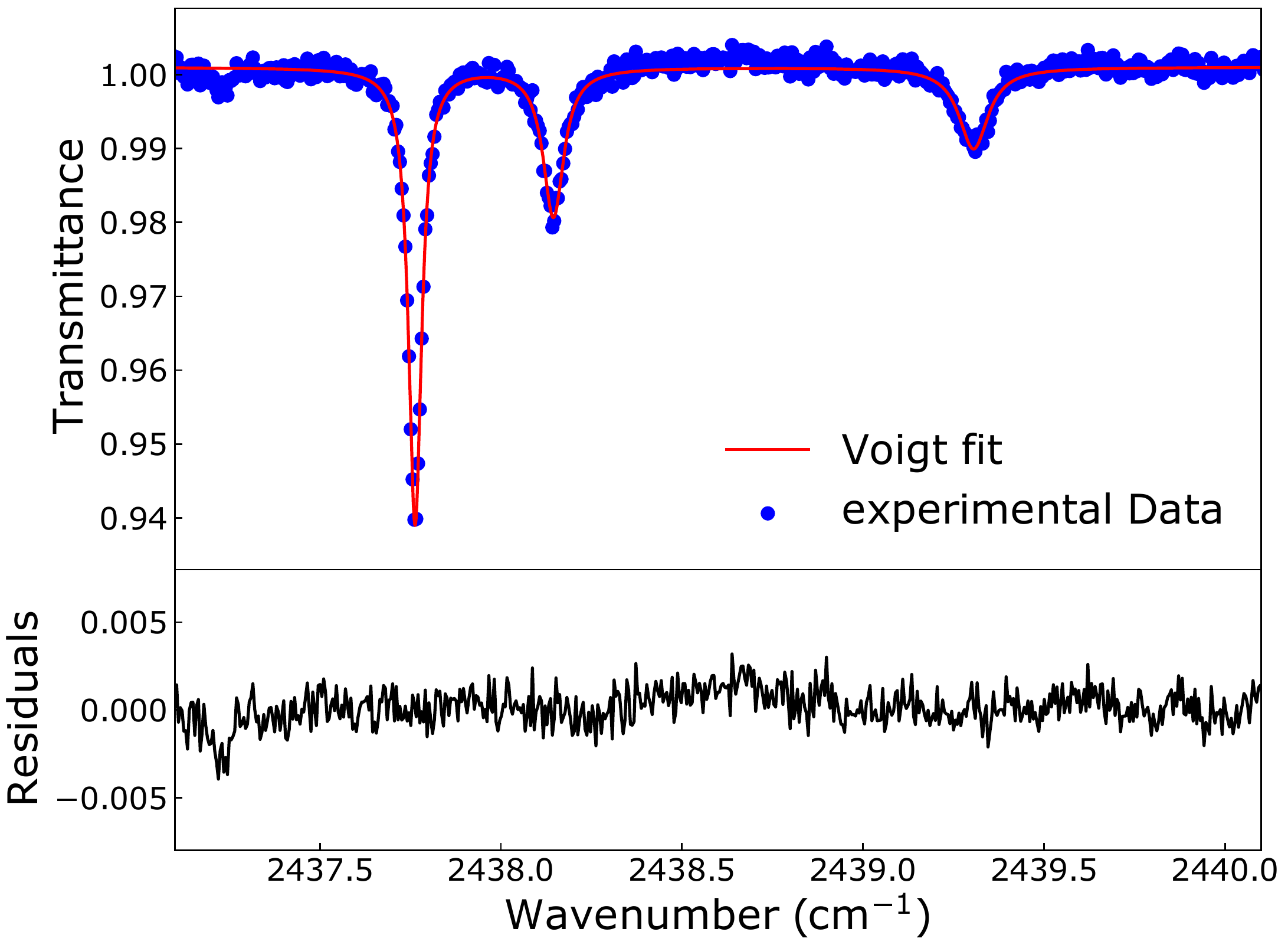}%
   \caption{Results of the non-linear least-squares fit of the $aR(2,K)$ multiplet assuming Voigt line profiles; $K=0\ldots2$
   correspond to the peaks from left to right. Blue dots depict the experimental spectrum, the red line depicts our multipeak Voigt
   fit. The lower panel shows the residual difference between experiment and simulation, the noise level is around 0.001.}
   \label{fig:voigt}
\end{figure}

For the fit we chose six typical peaks with high absorption, corresponding to the $aR(2,K)$ and $aR(3,K)$ multiplets in the
$3\nu_{2}$ band. In order to show the performance of the fit, we show the $aR(2,K)$ multiplet in \autoref{fig:voigt}, which fit
very well. The small dip at the low-wavenumber end of the residual difference corresponds to a $sP(6,1)$ transition in
$\nu_{2}+\nu_{4}$ band. From the fit, we obtained the line-transition frequencies $\nu_{ij}$, line intensities $I_{ij}$, and the
pressure-broadening component $\gamma_{p,t}$, specifically the self-broadening component $\gamma_\text{self}$ since the
gas cell only contained pure $^{14}${NH}$_{3}$.

\begin{figure}[h]
   \includegraphics[width=\linewidth]{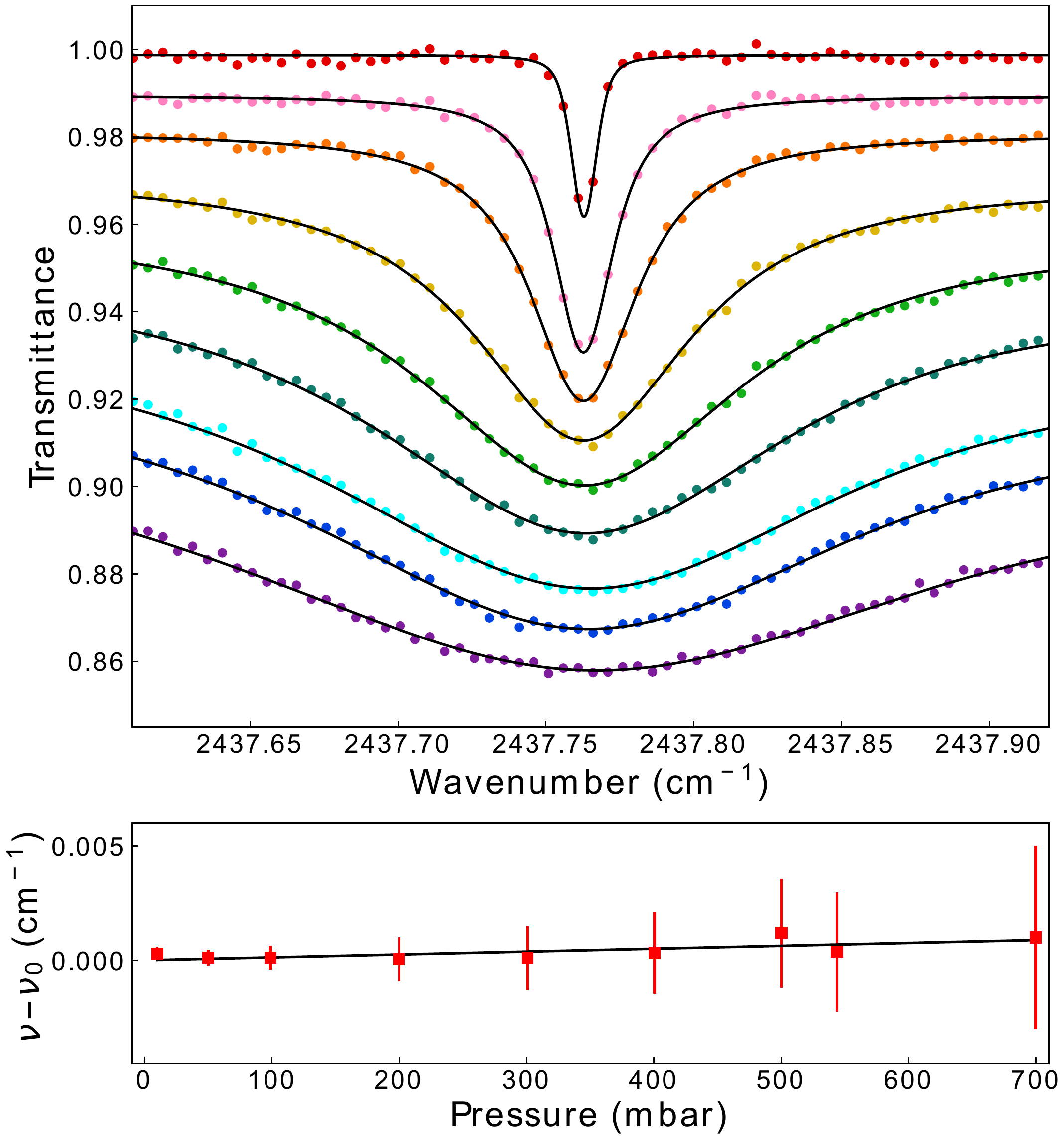}
   \caption{Extraction of the self-broadening and self-shift parameters from the lineshape of the $aR(2,0)$ transition at various
   pressures, \ie, from top to bottom 10.00~mbar, 50.00~mbar, 98.90~mbar, 199.90~mbar, 300.70~mbar, 400.50~mbar,
   500.01~mbar, 543.85~mbar, and 699.75~mbar, respectively. The black curves show the fitted Voigt profiles. For clarity, all
   lines are successively offset by -0.01 from top to bottom. The bottom panel shows the relative line transition wavenumbers
   (red squares) with respect to the vacuum line transition. Error bars depict the standard deviation from the multipeak Voigt fit.
   The black line is a linear fit of these transition frequencies from which the vacuum transition wavenumber and the self-shift are
   derived.}
   \label{fig:pressure}
\end{figure}

We recorded and averaged 25 spectra at each of the nine different pressures ranging from 10.00~mbar to 700.00~mbar. It took
around 1 minute for recording the spectra and pressures stayed the same during the 25 measurements. \autoref{fig:pressure}
shows the performance of the global multipeak Voigt fit and the extraction of self-broadening and self-shift coefficients for the
$aR(2,0)$ transition. For each pressure, $\nu_{ij}$, $I_{ij}$ and $\gamma_\text{self}$ were determined as described above. The
mean value of $I_{ij}$ and $\gamma_\text{self}$ was extracted. From a linear regression of the line transition wavenumbers
$\nu_{ij}$ as a function of pressure, see \autoref{fig:pressure}, the vacuum transition wavenumber $\nu_{ij}^0$ was determined
as the $y$ intercept and its self-shift parameter $\gamma_\text{self}$ as the slope $k$.

Line wavenumbers $\nu_{ij}^0$, line intensities $I_{ij}$, self-broadening $\gamma_\text{self}$, and self-shift $\delta_\text{self}$
components of the $aR(2,K)$ and $aR(3,K)$ multiplet in the $3\nu_{2}$ band of $^{14}${NH}$_{3}$ are presented in
\autoref{tab:results}. Specified uncertainties for line wavenumbers and self-shifts are their standard deviations from the fit, the
uncertainties for line intensities and self-broadenings were the combination of statistical errors from the averaging of multiple
scans and the standard deviations from the fit. Other systematic errors such as misalignments of the HeNe laser and MIR FC, the
line mixing, the FFT fliter etc.\ were neglected because they were relatively small in relation to the errors from the fit.
Line-transition wavenumbers and intensities show good agreement with the HITRAN database values~\cite{Gordon:JQSRT203:3}.
Our retrieved self-broadening parameters were around 20\% smaller than in the HITRAN database, which were derived from
measurements at 10~\um~\cite{Nemtchinov:JQSRT83:243}. Our results provide new laboratory data allowing to improve
pressure-self-broadening models.

\section{Conclusions}
\label{sec:conclusions}
We recorded the broadband absorption spectra of the $3\nu_{2} $ band of ammonia ($^{14}${NH}$_{3}$) near 4~\um using a
MIR-FC-based FTS system. Spectra were background corrected by FFT-filtering raw spectra for experimental etalon. A global
multipeak Voigt fit allowed us to determine transition wavenumbers and intensities, pressure self-broadening and self-shift
parameters at room temperature ($T=296$~K) for 6 prototypical $R$ branch rovibrational transitions.

The recorded line wavenumbers and line intensities are in good agreement with the HITRAN database. As no literature values
were available for the self-broadening and self-shift parameters, our new experimental data on these parameters provide useful
information about the molecular motions and collisions, which could help to improve their theoretical modeling.

Our experimental system is capable of trace-gas detection of different molecular species in the spectral range of $3\ldots6$~\um
with a current resolution of 0.00501~\invcm. We plan to implement a liquid-nitrogen-cooled multi-pass-cell cryostat to increase
the sensitivity.

\section{Acknowledgments}
We acknowledge support by Deutsches Elektronen-Synchrotron DESY, a member of the Helmholtz Association (HGF), and
through the Maxwell computational resources operated at Deutsches  Elektronen-Synchrotron DESY, Hamburg, Germany. This
work was further supported by the Deutsche Forschungsgemeinschaft (DFG) through the priority program ``Quantum Dynamics
in Tailored Intense Fields'' (QUTIF, SPP1840, YA~610/1) and the cluster of excellence ``Advanced Imaging of Matter'' (AIM,
EXC~2056, ID~390715994). G.Y.\ gratefully acknowledges financial support from the China Scholarship Council (CSC).

\pagebreak
\bibliography{string,cmi}

%\onecolumngrid
%\listofnotes
\end{document}